\documentclass[twocolumn,showpacs,preprintnumbers,amsmath,amssymb]{revtex4}
\usepackage{graphicx}

\begin{document}
\title{Upper bounds of a class of imperfect quantum sealing protocols}
\author{Guang-Ping He}
\affiliation{Department of Physics and Advanced Research Center,
Zhongshan University, Guangzhou 510275, China}

\begin{abstract}
The model of the quantum protocols sealing a classical bit is studied. It is
shown that there exist upper bounds on its security. For any protocol where
the bit can be read correctly with the probability $\alpha $, and reading
the bit can be detected with the probability $\beta $, the upper bounds are $%
\beta \leqslant 1/2$\ and $\alpha +\beta \leqslant 9/8$.
\end{abstract}

\pacs{03.67.Dd, 03.67.Hk, 03.67.-a, 89.70.+c}
\maketitle

\newpage

\section{Introduction}

Data sealing is a cryptographic problem between two parties. A sender
(Alice) stores some secret data in a certain form, so that any other reader
(Bob) can read it without Alice's helping. Meanwhile, if the data has been
read, it should be detectable by Alice\cite{note1}. A common example of
classical data sealing is closing a letter in an envelop with a wafer of
molten wax, into which was pressed the distinctive seal of the sender.

Like all other classical cryptographic protocols, it is interesting to find
the quantum version of data sealing for better security.
Bechmann-Pasquinucci \cite{sealing} first proposed a protocol which seals a
classical bit with a three-qubit state. Singh and Srikanth\cite{Srikanth}
extended the idea into a many-qubit majority voting scheme, and associated
it with secret sharing to improve the security. Chau\cite{seal q} presented
a protocol which seals quantum data with quantum error correcting code. The
protocol for sealing a classical string was also proposed\cite{He}.

However, as pointed out by the author himself, the protocol in Ref.\cite
{sealing} is insecure against collective measurements. More general, it was
further proven\cite{impossibility} that perfect quantum sealing of a
classical bit is impossible in principle. If a protocol allows the bit to be
perfectly retrievable by the reader, then collective measurements exist
which can read the bit without disturbing the corresponding quantum state.
It means that Bob can always read the bit without being detected by Alice.
In fact, the protocol for sealing quantum data cannot be used for sealing a
classical bit either. This is because the quantum states used for encoding
the bits $0$ and $1$ respectively are orthogonal to each other. They can be
distinguished by collective measurements without any disturbance too.

Therefore, it is natural to ask whether imperfect quantum sealing of a
classical bit is possible. Here ``imperfect'' means that the sealed bit $b$
is not perfectly retrievable in the protocol. Bob can only read $b$
correctly with the probability $\alpha <1$, while reading $b$ can be
detected with the probability $\beta $. Obviously a protocol with $\alpha =1$
is a perfect one. But if there exists a protocol in which both $\alpha $\
and $\beta $\ are less than but very close to $1$, it is still very valuable
for practical usage.

Nevertheless, in this paper it will be shown that upper bounds exist for $%
\alpha $\ and $\beta $. If Bob uses collective measurements instead of the
honest operations to read $b$, then the upper bounds are $\beta \leqslant
1/2 $\ and $\alpha +\beta \leqslant 9/8$. This result actually bounds the
power of practical quantum sealing of a classical bit. In the next section
we will establish a general model of imperfect quantum sealing protocols.
Basing on the model, the upper bounds will be obtained in Section III. In
Section IV some examples of imperfect protocols are studied. The impacts of
the result will be discussed in the last section.

\section{The Model}

First let us establish the model of imperfect quantum sealing protocols, on
which the discussion in this paper is based. Note that here and in the
following content, when speaking of quantum sealing protocols, we means the
protocols for sealing {\it a classical bit} only, except where noted. The
details of the data sealing process is not important to our discussion, but
the ending of the protocol generally has the following features:

(1) Bob knows an operation $P$;

(2) Bob owns a quantum system $\Psi $ ($\Psi $\ may not be in the eigenstate
of $P$. Otherwise the protocol becomes a perfect one);

(3) Alice lets Bob know two sets $G_{0}$, $G_{1}$ ($G_{0}\cap
G_{1}=\emptyset $), such that if he applies $P$ on $\Psi $\ and the outcome
is $g\in G_{0}$ ($g\in G_{1}$), he should take the value of the sealed bit
as $b^{\prime }=0$ ($b^{\prime }=1$); while if $g\notin G_{0}\cup G_{1}$,
the sealed bit cannot be identified, i.e. Bob needs to guess $b^{\prime }$
randomly by himself. Note that since $\Psi $\ may not be in the eigenstate
of $P$, the value of $b^{\prime }$ thus obtained will match Alice's input $b$
with a probability $\alpha $ only;

(4) Alice owns a quantum system $\Phi $ entangled with $\Psi $. And she
knows that the initial state of the system $\Phi \otimes \Psi $ is $\left|
\phi \otimes \psi \right\rangle $;

(5) At any time Alice can access to the entire system $\Phi \otimes \Psi $
and compare its state $\left| \phi ^{\prime }\otimes \psi ^{\prime
}\right\rangle $ with $\left| \phi \otimes \psi \right\rangle $. Thus she
can detect whether $b$ has been read with the probability $\beta =1-\left|
\left\langle \phi \otimes \psi \right| \left. \phi ^{\prime }\otimes \psi
^{\prime }\right\rangle \right| ^{2}$.

Note that in the protocols previously proposed (e.g. Refs.\cite
{sealing,Srikanth}), Alice does not own the system $\Phi $ described above,
and the case $g\notin G_{0}\cup G_{1}$ generally will not occur. But to make
our result as general as possible so that it may cover other protocols
potentially existed, we include these features in the model. Obviously
previous protocols are only the special cases\ of the model where the
entanglement between $\Phi $ and $\Psi $ has already collapsed before the
end of the protocol, and $G_{0}$, $G_{1}$ cover all possible outcomes of $g$.

\section{The Upper Bounds}

Let $H$\ be the global Hilbert space constructed by all possible states of $%
\Psi $. $\{\left| \hat{e}_{i}\right\rangle \}$ denotes a basis of $H$, which
is the eigenvector set of the operation $P$. It can be divided into three
orthogonal subsets $\{\left| \hat{e}_{i}^{(0)}\right\rangle \}$, $\{\left|
\hat{e}_{i}^{(1)}\right\rangle \}$\ and $\{\left| \hat{e}_{i}^{(2)}\right%
\rangle \}$, such that $\forall \left| \psi \right\rangle \in \{\left|
\hat{e}_{i}^{(0)}\right\rangle \}$ ($\forall \left| \psi \right\rangle \in
\{\left| \hat{e}_{i}^{(1)}\right\rangle \}$), applying $P$ on $\left| \psi
\right\rangle $\ will lead to $b^{\prime }=0$ ($b^{\prime }=1$); and $%
\forall \left| \psi \right\rangle \in \{\left| \hat{e}_{i}^{(2)}\right%
\rangle \}$\ will lead to $b^{\prime }=0$ and $b^{\prime }=1$ with the equal
probability $1/2$. $H$ is consequently divided into three subspaces $H_{0}$,
$H_{1}$ and $H_{2}$, whose eigenvector sets are $\{\left| \hat{e}%
_{i}^{(0)}\right\rangle \}$, $\{\left| \hat{e}_{i}^{(1)}\right\rangle \}$\
and $\{\left| \hat{e}_{i}^{(2)}\right\rangle \}$ respectively.

Though the subspaces $H_{0}$, $H_{1}$ and $H_{2}$ are orthogonal to each
other, the quantum states used for encoding the bits $0$ and $1$
respectively need not to be orthogonal in an imperfect quantum sealing
protocol. Instead, any state in $H$ could be the state of the system $\Psi $%
. That is, the state of $\Psi $ may contain the vectors from different
subspaces, so that the states used for encoding $0$ and $1$ may overlap. In
this case, the cheating strategy in the impossibility proof of perfect
quantum sealing can not be successful with the probability $1$.

However, note that $\{\left| \hat{e}_{i}^{(0)}\right\rangle \}$, $\{\left|
\hat{e}_{i}^{(1)}\right\rangle \}$\ and $\{\left| \hat{e}_{i}^{(2)}\right%
\rangle \}$ form a complete basis of $H$. Let $\left| \phi _{0}\otimes \psi
_{0}\right\rangle $ ($\left| \phi _{1}\otimes \psi _{1}\right\rangle $)
denotes the initial state of $\Phi \otimes \Psi $ when Alice want to seal $%
b=0$ ($b=1$). It can always be expanded as
\begin{eqnarray}
\left| \phi _{b}\otimes \psi _{b}\right\rangle  &=&\sqrt{\alpha _{b}^{(0)}}%
\sum\limits_{i}\sqrt{\lambda _{b,i}^{(0)}}\left| \hat{f}_{i}^{(0)}\right%
\rangle \left| \hat{e}_{i}^{(0)}\right\rangle   \nonumber \\
&&+\sqrt{\alpha _{b}^{(1)}}\sum\limits_{i}\sqrt{\lambda _{b,i}^{(1)}}\left|
\hat{f}_{i}^{(1)}\right\rangle \left| \hat{e}_{i}^{(1)}\right\rangle
\nonumber \\
&&+\sqrt{\alpha _{b}^{(2)}}\sum\limits_{i}\sqrt{\lambda _{b,i}^{(2)}}\left|
\hat{f}_{i}^{(2)}\right\rangle \left| \hat{e}_{i}^{(2)}\right\rangle ,
\label{psai}
\end{eqnarray}
with $\alpha _{b}^{(0)}+\alpha _{b}^{(1)}+\alpha _{b}^{(2)}=1$, $%
\sum\limits_{i}\lambda _{b,i}^{(0)}=\sum\limits_{i}\lambda
_{b,i}^{(1)}=\sum\limits_{i}\lambda _{b,i}^{(2)}=1$\ (sum over all possible $%
i$ within each corresponding subspace), $b=0,1$. All $\left| \hat{f}%
_{i}\right\rangle $\ are the vectors describing the state of $\Phi $, which
are not required to be orthogonal to each other.

With such initial states, it can be seen that the maximal probability for
Bob to read $b$ correctly (i.e. his outcome $b^{\prime }$ matches Alice's
input $b$) is
\begin{equation}
\alpha =[(\alpha _{0}^{(0)}+\alpha _{0}^{(2)}/2)+(\alpha _{1}^{(1)}+\alpha
_{1}^{(2)}/2)]/2.  \label{alfa}
\end{equation}

Consider a dishonest Bob who does not use $P$ but the following strategy to
read $b$. Define two operators $P_{j}=\sum\limits_{i}\left| \hat{e}%
_{i}^{(j)}\right\rangle \left\langle \hat{e}_{i}^{(j)}\right| $\ (sum over
all possible $i$ within $H_{j}$. $j=0,1$), which are the collective
measurements projecting the quantum state into the subspaces $H_{0}$ and $%
H_{1}$ respectively. Bob randomly chooses to apply either $P_{0}$ or $P_{1}$
on $\Psi $ with the equal probability $1/2$. If his choice is $P_{j}$ and $%
\Psi $ can be projected to the subspace $H_{j}$ successfully, he takes $%
b^{\prime }=j$; else if the projection fails, he always takes $b^{\prime }=%
\bar{j}$ without further measurements on the quantum state. That is, he
never makes further attempts to distinguish the state between the two
subspaces $H_{\bar{j}}$\ and $H_{2}$. It can be shown that with this
strategy, Bob can also reach the maximized $\alpha $ in Eq.(\ref{alfa}). Now
let us calculate the probability $\beta $ for the reading to be detected by
Alice. There can be four different cases:

(1) {\it Bob's choice is} $P_{0}${\it , and the initial state of} $\Phi
\otimes \Psi $ {\it is} $\left| \phi \otimes \psi \right\rangle =\left| \phi
_{0}\otimes \psi _{0}\right\rangle ${\it .} From Eq.(\ref{psai}) it can be
seen that $\Psi $ can be projected into $H_{0}$\ successfully with the
probability $\alpha _{0}^{(0)}$. Meanwhile, $\Phi \otimes \Psi $ collapses
to $\left| \phi ^{\prime }\otimes \psi ^{\prime }\right\rangle
=\sum\limits_{i}\sqrt{\lambda _{0,i}^{(0)}}\left| \hat{f}_{i}^{(0)}\right%
\rangle \left| \hat{e}_{i}^{(0)}\right\rangle $. It can be viewed as the
initial state with the probability $p_{0s}=\left| \left\langle \phi \otimes
\psi \right| \left. \phi ^{\prime }\otimes \psi ^{\prime }\right\rangle
\right| ^{2}=\alpha _{0}^{(0)}$. Therefore Alice can detect the disturbance
of the state with the probability 
\begin{equation}
\beta _{0s}=1-p_{0s}=1-\alpha _{0}^{(0)}.
\end{equation}

On the other hand, the projection can also fail with the probability $\alpha
_{0}^{(1)}+\alpha _{0}^{(2)}$, with $\Phi \otimes \Psi $ collapsing to $%
\left| \phi ^{\prime }\otimes \psi ^{\prime }\right\rangle =(\sqrt{\alpha
_{0}^{(1)}}\sum\limits_{i}\sqrt{\lambda _{0,i}^{(1)}}\left| \hat{f}%
_{i}^{(1)}\right\rangle \left| \hat{e}_{i}^{(1)}\right\rangle +\sqrt{\alpha
_{0}^{(2)}}\sum\limits_{i}\sqrt{\lambda _{0,i}^{(2)}}\left| \hat{f}%
_{i}^{(2)}\right\rangle \left| \hat{e}_{i}^{(2)}\right\rangle )/\sqrt{\alpha
_{0}^{(1)}+\alpha _{0}^{(2)}}$. It can be viewed as the initial state with
the probability $p_{0f}=\alpha _{0}^{(1)}+\alpha _{0}^{(2)}$. Therefore
Alice can detect the disturbance with the probability 
\begin{equation}
\beta _{0f}=1-p_{0f}=1-\alpha _{0}^{(1)}-\alpha _{0}^{(2)}.
\end{equation}

Altogether, in this case reading $b$ can be detected by Alice with the
probability 
\begin{eqnarray}
\beta _{0} &=&\alpha _{0}^{(0)}\beta _{0s}+(\alpha _{0}^{(1)}+\alpha
_{0}^{(2)})\beta _{0f}  \nonumber \\
&=&2\alpha _{0}^{(0)}(1-\alpha _{0}^{(0)}).  \label{case1}
\end{eqnarray}
Here the condition $\alpha _{0}^{(0)}+\alpha _{0}^{(1)}+\alpha _{0}^{(2)}=1$
is used.

(2) {\it Bob's choice is} $P_{0}${\it , and the initial state of} $\Phi
\otimes \Psi $ {\it is} $\left| \phi \otimes \psi \right\rangle =\left| \phi
_{1}\otimes \psi _{1}\right\rangle ${\it .} Similar to the analysis above,
in this case reading $b$ can be detected by Alice with the probability 
\begin{equation}
\beta _{1}=2\alpha _{1}^{(0)}(1-\alpha _{1}^{(0)}).  \label{case2}
\end{equation}

(3) {\it Bob's choice is} $P_{1}${\it , and the initial state is} $\left|
\phi \otimes \psi \right\rangle =\left| \phi _{0}\otimes \psi
_{0}\right\rangle ${\it .} Similar to Eq.(\ref{case1}), the detecting
probability is 
\begin{equation}
\beta _{0}^{\prime }=2\alpha _{0}^{(1)}(1-\alpha _{0}^{(1)}).
\end{equation}

(4) {\it Bob's choice is} $P_{1}${\it , and the initial state is} $\left|
\phi \otimes \psi \right\rangle =\left| \phi _{1}\otimes \psi
_{1}\right\rangle ${\it .} Similar to Eq.(\ref{case2}), the detecting
probability is 
\begin{equation}
\beta _{1}^{\prime }=2\alpha _{1}^{(1)}(1-\alpha _{1}^{(1)}).
\end{equation}

In all, the average probability for Alice to detect the reading is 
\begin{eqnarray}
\beta &=&[\alpha _{0}^{(0)}(1-\alpha _{0}^{(0)})+\alpha _{1}^{(0)}(1-\alpha
_{1}^{(0)})  \nonumber \\
&&+\alpha _{0}^{(1)}(1-\alpha _{0}^{(1)})+\alpha _{1}^{(1)}(1-\alpha
_{1}^{(1)})]/2.  \label{beta}
\end{eqnarray}
The right-hand side of this equation reaches its maximum when $\alpha
_{0}^{(0)}=\alpha _{1}^{(0)}=\alpha _{0}^{(1)}=\alpha _{1}^{(1)}=1/2$ and $%
\alpha _{0}^{(2)}=\alpha _{0}^{(2)}=0$. Thus the upper bound for $\beta $ is 
\begin{equation}
\beta \leqslant 1/2.  \label{bound1}
\end{equation}

Combining Eqs.(\ref{alfa}) and (\ref{beta}), we have 
\begin{eqnarray}
\alpha +\beta &=&[\alpha _{0}^{(0)}(3/2-\alpha _{0}^{(0)})+\alpha
_{0}^{(1)}(1/2-\alpha _{0}^{(1)})  \nonumber \\
&&+\alpha _{1}^{(1)}(3/2-\alpha _{1}^{(1)})+\alpha _{1}^{(0)}(1/2-\alpha
_{1}^{(0)})  \nonumber \\
&&+1]/2.  \label{sum}
\end{eqnarray}
The right-hand side of this equation reaches its maximum when $\alpha
_{0}^{(0)}=\alpha _{1}^{(1)}=3/4$, $\alpha _{0}^{(1)}=\alpha _{1}^{(0)}=1/4$
and $\alpha _{0}^{(2)}=\alpha _{0}^{(2)}=0$. Thus we obtain another upper
bound 
\begin{equation}
\alpha +\beta \leqslant 9/8.  \label{bound2}
\end{equation}

\section{Examples}

\subsection{Breaking the majority voting scheme}

Here we give an example on how the above cheating strategy is applied to an
embodied protocol.

In Ref.\cite{Srikanth} a compound scheme was proposed, which embeds a
majority voting quantum sealing scheme in a classical secret sharing scheme.
As pointed out by the authors, the entire scheme is secure since the secret
sharing scheme prevents any single reader from possessing all the qubits to
perform collective measurements. Without the secret sharing scheme, as it
will be shown below, the majority voting scheme alone is insecure.

The majority voting scheme is defined as follows. To seal a classical bit $b$%
, Alice prepares $n$ qubits. $fn$ ($f<1/2$) qubits called the code qubits
are prepared in the state $\left| b\right\rangle $. The other $(1-f)n$\
qubits called the seal qubits are put randomly in any eigenstate of the
diagonal basis $\left| \pm \right\rangle \equiv (\left| 0\right\rangle \pm
\left| 1\right\rangle )/\sqrt{2}$. To read the bit $b$, an honest Bob is
supposed to measure each qubit in the computational basis $\{\left|
0\right\rangle ,\left| 1\right\rangle \}$. He takes $b=0$ (or $1$) if more
than $n/2$ qubits are found as $\left| 0\right\rangle $\ (or $\left|
1\right\rangle $). Since $f<1/2$, we can see that the states encoding $0$
and $1$ are nonorthogonal to each other. Bob stands a nonzero probability of
misreading the bit. Therefore it is an imperfect sealing scheme.

The cheating strategy to this scheme is exactly the one described in the
previous section. Let $w(c)$ denote the weight (the number of the bit $1$)
of a classical binary $n$-bit string $c$. The eigenvector set $\{\left| 
\hat{e}_{i}\right\rangle \}$\ in the present case is the computational basis 
$\{\left| c\right\rangle \}$ of the global Hilbert space $H$, where $c$\
runs through all possible classical $n$-bit strings. Its three orthogonal
subsets are defined as $\{\left| \hat{e}_{i}^{(0)}\right\rangle \}\equiv
\{\left| c\right\rangle |w(c)<n/2\}$, $\{\left| \hat{e}_{i}^{(1)}\right%
\rangle \}\equiv \{\left| c\right\rangle |w(c)>n/2\}$\ and $\{\left| \hat{e}%
_{i}^{(2)}\right\rangle \}\equiv \{\left| c\right\rangle |w(c)=n/2\}$. The
three subspaces $H_{0}$, $H_{1}$ and $H_{2}$ are consequently defined by
these subsets.

For simplicity let us consider the case where $n$ is odd so that $\{\left|
\hat{e}_{i}^{(2)}\right\rangle \}=\emptyset $. Let $\left| \psi
_{b}\right\rangle $\ be the state of the $n$-qubit system encoding $b$. For
any $\left| \psi _{b}\right\rangle $ with $fn$ code qubits, we can expand
all the $(1-f)n$ seal qubits in the computational basis. Thus $\left| \psi
_{b}\right\rangle $ is expanded into $N\equiv 2^{(1-f)n}$ items, of which $%
M^{(\bar{b})}\equiv \sum_{i=(n+1)/2}^{(1-f)n}%
{(1-f)n \choose i}%
$ items belong to $\{\left| \hat{e}_{i}^{(\bar{b})}\right\rangle \}$, while
the other $M^{(b)}\equiv N-M^{(\bar{b})}$ items belong to $\{\left| \hat{e}%
_{i}^{(b)}\right\rangle \}$. That is
\begin{eqnarray}
\left| \psi _{b}\right\rangle  &=&\sqrt{M^{(b)}/N}\sum\limits_{i}(\Lambda
_{i}^{(b)}/\sqrt{M^{(b)}})\left| \hat{e}_{i}^{(b)}\right\rangle   \nonumber
\\
&&+\sqrt{M^{(\bar{b})}/N}\sum\limits_{i}(\Lambda _{i}^{(\bar{b})}/\sqrt{M^{(%
\bar{b})}})\left| \hat{e}_{i}^{(\bar{b})}\right\rangle ,  \label{ex1}
\end{eqnarray}
where any $\Lambda _{i}$\ can only be $\pm 1$\ or $0$, with $%
\sum\limits_{i}\Lambda _{i}^{(b)}=\sqrt{M^{(b)}}$, $\sum\limits_{i}\Lambda
_{i}^{(\bar{b})}=\sqrt{M^{(\bar{b})}}$. For example, an $n=5$, $f=0.4$ state
$\left| 11+-+\right\rangle $\ sealing $b=1$ (the first one in Eq.(1) of Ref.
\cite{Srikanth}) can be expanded as
\begin{eqnarray}
\left| 11+-+\right\rangle  &=&(\left| 11000\right\rangle   \nonumber \\
&&+\left| 11001\right\rangle -\left| 11010\right\rangle -\left|
11011\right\rangle   \nonumber \\
&&+\left| 11100\right\rangle +\left| 11101\right\rangle -\left|
11110\right\rangle   \nonumber \\
&&-\left| 11111\right\rangle )/\sqrt{8},
\end{eqnarray}
where the first item on the right of the equation belongs to $\{\left|
\hat{e}_{i}^{(0)}\right\rangle \}$, while the other $7$ items belong to $%
\{\left| \hat{e}_{i}^{(1)}\right\rangle \}$. Obviously when $\left| \psi
_{b}\right\rangle $ is measured honestly, $b$ can be read correctly with
the probability
\begin{equation}
\alpha =M^{(b)}/N=1-\sum_{i=(n+1)/2}^{(1-f)n}%
{(1-f)n \choose i}%
/2^{(1-f)n}.  \label{exalfa}
\end{equation}

On the other hand, a dishonest Bob can always use the operator $P_{0}=\sum
\left| c\right\rangle \left\langle c\right| $\ (sum over all $c$ satisfying $%
w(c)<n/2$) to perform a collective measurement on $\left| \psi
_{b}\right\rangle $. He takes $b=0$ whenever the projection is successful,
else he takes $b=1$. It can be seen that he can also read $b$ correctly with
the above probability $\alpha $. The probability $\beta $\ for the reading
to be detected by Alice can be calculated by repeating the analysis in the
previous section. By comparing Eq.(\ref{ex1}) with Eq.(\ref{psai}), we find $%
\alpha _{b}^{(b)}=M^{(b)}/N$, $\alpha _{b}^{(\bar{b})}=M^{(\bar{b})}/N$ and $%
\alpha _{b}^{(2)}=0$. Substituting them into Eq.(\ref{beta}) gives

\begin{eqnarray}
\beta  &=&\alpha _{b}^{(b)}(1-\alpha _{b}^{(b)})+\alpha _{b}^{(\bar{b}%
)}(1-\alpha _{b}^{(\bar{b})})  \nonumber \\
&=&2\alpha (1-\alpha )  \nonumber \\
&=&2\left( 1-\sum_{i=(n+1)/2}^{(1-f)n}%
{(1-f)n \choose i}%
/2^{(1-f)n}\right)   \nonumber \\
&&\cdot \sum_{i=(n+1)/2}^{(1-f)n}%
{(1-f)n \choose i}%
/2^{(1-f)n}.  \label{exbeta}
\end{eqnarray}

Eqs.(\ref{exalfa}) and (\ref{exbeta}) clearly show that in the majority
voting scheme, if Alice wants to rise the readability $\alpha $, the
probability $\beta $ for her to detect the reading will inevitably drop no
matter how she chooses $n$ and $f$. Especially, $\beta \rightarrow 0$ when $%
\alpha \rightarrow 1$. This result as well as the value of $\alpha +\beta $
as a function of $\alpha $ are plotted in Fig.1. Thus we see that without
being associated with secret sharing, the majority voting scheme alone is
insecure against the collective measurement.

\begin{figure}
\includegraphics{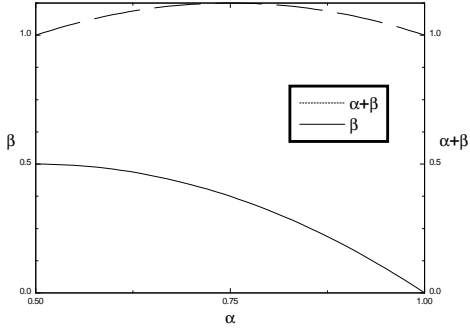}
\caption{\label{fig:epsart}The security of the majority voting
scheme. $\alpha $ is the probability for Bob to read the bit
successfully. $\beta $ is the
probability for Alice to detect the reading. The solid line represents $%
\beta $ as a function of $\alpha $. The dashed line represents
$\alpha +\beta $ as a function of $\alpha $.}
\end{figure}

\subsection{The scheme that reaches the upper bound $\protect\beta =1/2$}

Eq.(\ref{beta}) indicates that to reach the maximum $\beta =1/2$, Alice
should seal the bit $b$ in the form
\begin{eqnarray}
\left| \phi _{b}\otimes \psi _{b}\right\rangle &=&(\sum\limits_{i}\sqrt{%
\lambda _{b,i}^{(0)}}\left| \hat{f}_{i}^{(0)}\right\rangle \left| \hat{e}%
_{i}^{(0)}\right\rangle  \nonumber \\
&&+\sum\limits_{i}\sqrt{\lambda _{b,i}^{(1)}}\left| \hat{f}%
_{i}^{(1)}\right\rangle \left| \hat{e}_{i}^{(1)}\right\rangle )/\sqrt{2.}
\end{eqnarray}
But clearly such a protocol is useless, since the bit can only be read
correctly with the probability $\alpha =1/2$. Even random guess based on
nothing at all can reach such a probability.

\subsection{The scheme that reaches the upper bound $\protect\alpha +\protect%
\beta =9/8$}

Eq.(\ref{sum}) indicates that to reach the maximum $\alpha +\beta =9/8$,
Alice should seal the bit $b$ in the form
\begin{eqnarray}
\left| \phi _{b}\otimes \psi _{b}\right\rangle  &=&\frac{\sqrt{3}}{2}%
\sum\limits_{i}\sqrt{\lambda _{b,i}^{(b)}}\left| \hat{f}_{i}^{(b)}\right%
\rangle \left| \hat{e}_{i}^{(b)}\right\rangle   \nonumber \\
&&+\frac{1}{2}\sum\limits_{i}\sqrt{\lambda _{b,i}^{(\bar{b})}}\left| \hat{f}%
_{i}^{(\bar{b})}\right\rangle \left| \hat{e}_{i}^{(\bar{b})}\right\rangle .
\end{eqnarray}
In this case $\alpha =3/4$, $\beta =3/8$. Note that keeping the system $\Phi
$ (whose state is described by $\left| \hat{f}_{i}\right\rangle $) at
Alice's side is important. Otherwise, consider the simplified version
\begin{equation}
\left| \psi _{b}\right\rangle =\frac{\sqrt{3}}{2}\sum\limits_{i}\sqrt{%
\lambda _{b,i}^{(b)}}\left| \hat{e}_{i}^{(b)}\right\rangle +\frac{1}{2}%
\sum\limits_{i}\sqrt{\lambda _{b,i}^{(\bar{b})}}\left| \hat{e}_{i}^{(\bar{b}%
)}\right\rangle .
\end{equation}
It cannot reach $\alpha +\beta =9/8$ if Bob knows that Alice has prepared
the states this way. This is because the cheating strategy in the previous
section does not require Bob to know the value of $\alpha $. If he does, he
may have other optimal strategies to further reduce $\beta $. In the present
case, Bob can fake the quantum state with a certain probability after
reading it. For example, if he has applied $P_{0}$ on $\Psi $\ and the
projection fails, he knows that the state has collapsed to $\sum\limits_{i}%
\sqrt{\lambda _{b,i}^{(1)}}\left| \hat{e}_{i}^{(1)}\right\rangle $. Since he
knows $\alpha $, he can use a unitary transformation to shift the state into 
$\sqrt{1-\alpha }\sum\limits_{i}\sqrt{1/N_{0}}\left| \hat{e}%
_{i}^{(0)}\right\rangle +\sqrt{\alpha }\sum\limits_{i}\sqrt{\lambda
_{b,i}^{(1)}}\left| \hat{e}_{i}^{(1)}\right\rangle $ ($N_{0}$ is the
dimensionality of $H_{0}$). This generally increases his chance to survive
through Alice's detection. Therefore the value of $\alpha +\beta $ in this
protocol will be further reduced. This is also true for the majority voting
scheme discussed in the section IV.A. If Bob knows $f$ he knows $\alpha $
from Eq.(\ref{exalfa}), so that he can further reduce $\beta $. But if there
is a system $\Phi $\ at Alice's side, it cannot be faked by Bob. Then
changing the state at his own side alone will be useless and the upperbound
could be reached.

\section{Discussion and Conclusion}

The upper bounds $\beta \leqslant 1/2$\ and $\alpha +\beta \leqslant 9/8$
found in this paper can be seen as an extension of the impossibility proof
of perfect quantum sealing of a classical bit\cite{impossibility}. The
latter can be seen as the special case where $\alpha =1$. Eq.(\ref{alfa})
shows that $\alpha =1$\ means $\alpha _{0}^{(0)}=\alpha _{1}^{(1)}=1$ and $%
\alpha _{0}^{(1)}=\alpha _{0}^{(2)}=\alpha _{1}^{(0)}=\alpha _{1}^{(2)}=0$.
Substituting these values into Eq.(\ref{beta}) immediately gives $\beta =0$.

Due to the existence of these upper bounds, quantum sealing seems to be far
from practical usage, unless we can find a protocol that cannot be covered
by the model proposed in Section II. However, so far there still has no sign
on the existence of such a protocol. Luckily, as mentioned above, our model
is limited to the protocols which seal a single classical bit. The protocol
which seals a classical string\cite{He} is not covered and can be secure.

Very recently, the insecurity of quantum sealing was also studied by
Chau\cite{Chau}, in which a more detailed model of quantum sealing ($\{g\notin
G_{0}\cup G_{1}\}=\emptyset $ and Bob knows $\alpha $) was analyzed with a
different approach. The result $\beta \leqslant 1/2$\ was also obtained,
which consists with the finding in the present paper.

I would like to thank Helle Bechmann-Pasquinucci, R. Srikanth and Hoi-Fung
Chau for their useful discussions.


\begin{thebibliography}{99}
\bibitem{note1}  In fact, in the original proposal of Ref.\cite{sealing},
the detection was suggested to be performed by another verifier authorized
by Alice. For simplicity without loss of generality, we may assume that the
verifier can get as much information as he needs from Alice, so that he can
maximize the successful probability of the detection. Therefore the verifier
can be taken as Alice herself.
\end{thebibliography}
\end{document}